# Modeling and Simulation of Electromigration Behavior for Via Array Structure

Karthik Airani, Rohit Guttal

Department of Silicon Engineering, Micron Technology

8000 South Federal Way, Boise, Idaho

**Keywords — electromigration, via array, uneven current distribution, stress time translation, cumulative failure distribution**

## Abstract

**In this paper, we develop a analytical model and algorithm for calculating uneven current distribution in via array structures. We propose a stress time translation formula and cumulative failure distribution equation to model the memory effect of electro migration stress or damage on a via array structure. We develop a method to project via array electromigration (EM) lifetime based on an arbitrary via failure sequence, and demonstrate that the proposed via array EM lifetime distribution trend correlates well with experimental results.**

## 1. Introduction

Electromigration (EM) is a phenomenon of material transport caused by a gradual movement of ions in a conductor due to momentum transfer between conducting electrons and diffusing metal atoms [1]. A void or an extrusion may be formed if a wire undergoes EM stress for a sufficiently long period of time. Such defects may eventually damage the wire. Today current density is increasing sharply and wires becoming narrower ever. All these trends are making it extremely difficult to close EM reliability for modern digital integrated circuit design.

Although people refer to wire EM, in reality, vias are the most sensitive to EM, and most of the EM defects are associated with vias. To overcome this restriction, designers have been widely using via array to strengthen its resistance to EM damage. However, even EM failure of a single via is well understood, it is not clear how via array behaves under EM stress. We found that via array has a very different EM characterteristics, therefore a more systematic study is needed.

One fundamental difference in via array is that current is not evenly distributed into each via in a via array. Thus, instead of using a lumped single via model, we need a finer model that can reflect this uneven current distribution [2]. Also, in most of the real designs, current flowing across via or via arrays are asymmetric, therefore magnifying the uneven current distribution effect. In our experiments, we first build a finite-element-method (FEM) in ANSYS to simulate the current flow in via array. In Figure 1 (b), we show FEM simulation results for a 2 x 2 via array with current values indicated in Figure 1 (a).

In Figure 1 (a), we show a typical via array structure, where two orthogonal wires are connected using a regular array of evenly spaced vias. To our best knowlede, no quantitative analysis of via array current distribution exists in literature, which motivates us to develop a model for fast calculation of current distribution in a via array and explain the cause of the observed uneven current distribution.

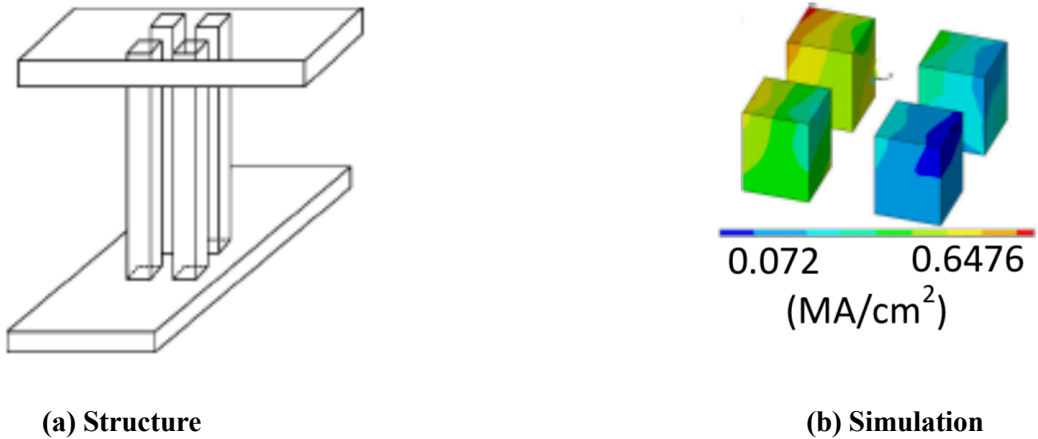

**(a) Structure**                                    **(b) Simulation**

**Figure 1. FEM Structure and simulation**

The redundancy of a via array is another important property that significantly impact EM reliability. As shown in Figure 2, the overall EM reliability of a double cut via with current density j is not equivalent to a single via with current density j.

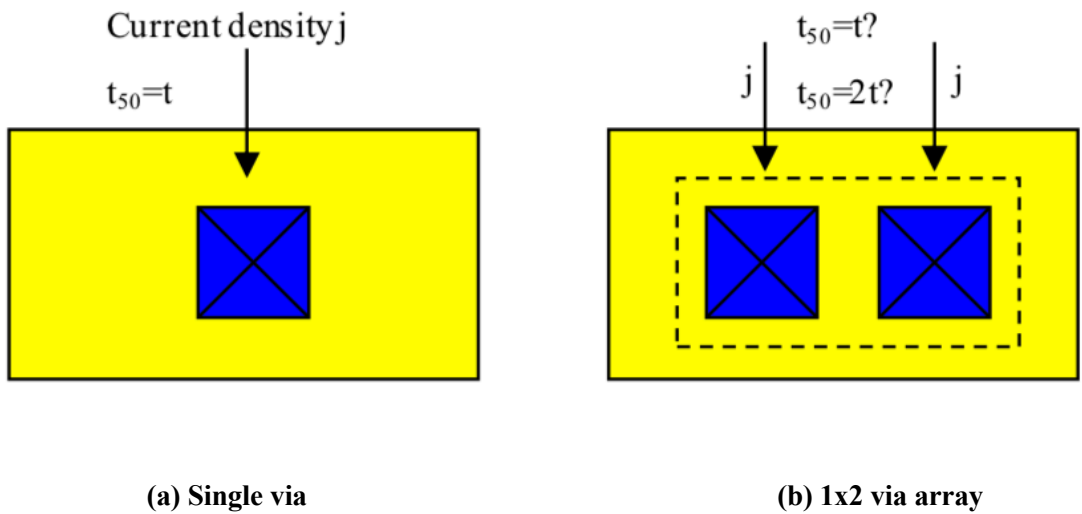

**(a) Single via**                                    **(b) 1x2 via array**

**Figure 2. Single via VS via array**

One question to ask is that assuming mean time to failure (MTTF or $t_{50}$) is t for Figure 2(a), what is $t_{50}$ in Figure 2(b)? To answer this question, we use a random failure process with lognormal distribution in this paper and explain how to generate the via failure sequences with Monte Carlo simulation to estimate the EM reliability of a via array.

## 2. Non-uniform Current Distribution

### A. FEM and Spice Simulation

We first build a finite element method (FEM) model to simulate and obtain via current distribution of a via array [3]. The experimental results show that a linear dependency of via current on their wire currents. Therefore, a much simpler resistive SPICE model is used to replace the FEM model. Figure 3 shows an example of SPICE mesh.

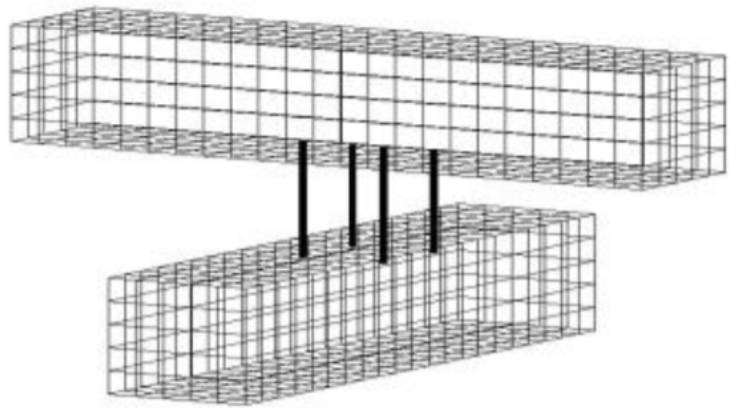

**Figure 3. 3D grid model used in SPICE simulations**

The simplified model captures only resistive effect, but when we compare the simulation results, the differences are well kept under 1%.

### B. Fast Via Current Computation Formula

We already see that using a simplified SPICE model can obtain current distribution much faster without accuracy loss. However, if an analytical formula can be derived, it will even faster than a SPICE mesh [4]. We use a 4x4 array as an example, the proposed approach can be generally expanded to an arbitrary via array.

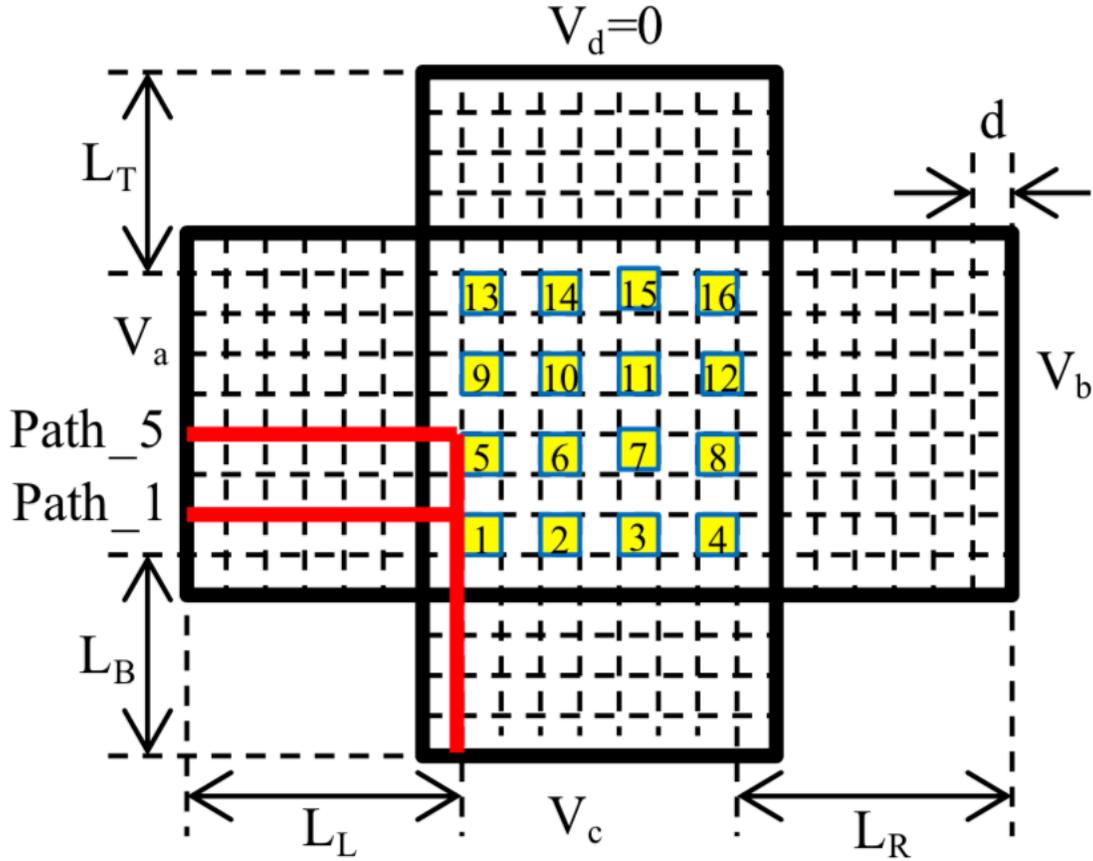

**Figure 4. 4x4 via array**

Consider two orthogonal wires *meatl 1* and *metal 2* connected through a 4x4 via array as in Figure 4(a). Vias are placed in the center of the wire intersection. The lengths of wires extended from the intersection area are $L_T$, $L_B$, $L_L$ and $L_R$ (assume $L_T = L_B$ and $L_L = L_R$) and their resistances are $R_T$, $R_B$, $R_L$ and $R_R$. The unit mesh gird is of length $d$ and resistance $r$. Voltages $V_a$, $V_b$, $V_c$ and $V_d$ are applied at the wire ends. $V_d$ is set to be the reference voltage ($V_d$=0). Our objective is to compute current flowing through each via, given the wire dimensions, applied voltage and via resistance $R_{via}$.

Using superposition property of the resistive circuit, we solve the system each time with only one voltage source present and all others shorted to ground and take summation of the results. Due to symmetry of the system, when only $V_a$ is applied, $I_{1\_a\_via}$=$I_{13\_a\_via}$, $I_{2\_a\_via}$=$I_{14\_a\_via}$, etc., where $I_{i\_a\_via}$ denotes the current flowing through via $i$ when $V_a$ is applied. The currents $I_{1\_a\_via}$, $I_{2\_a\_via}$, $I_{3\_a\_via}$ and $I_{4\_a\_via}$ are different due to the voltage gradient between left and right end of the horizontal wire. This is the forward voltage gradient of $V_a$. The currents $I_{1\_a\_via}$, $I_{5\_a\_via}$, are different due to the voltage gradient between left end of horizontal wire and bottom end of vertical wire. This is the lateral voltage gradient of $V_a$.

The forward gradient is defined as $f = V_{1a}/V_{2a}$, where $V_{ia}$ denotes voltage of top surface of via $i$, when $V_a$ is applied. We compute $f$ using resistance divider shown in Figure 5. The values of $R$, $R_{left}$, $R_{right}$ and $R_1$ can be extracted from the wire dimensions, grid branch resistance $r$ and via positions. $R_{via}+R$ is the equivalent resistance from a via to ground and $\bar{R} = R_B // R_T$.

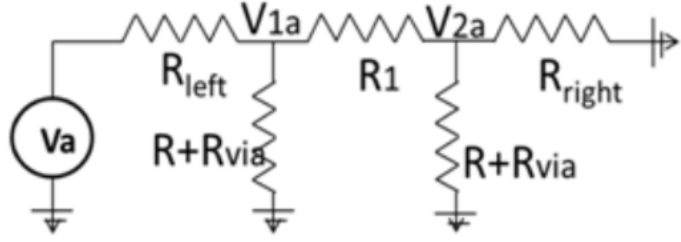

**Figure 5. Equivalent circuit for computing forward gradient**

For the configuration in Figure 5, we have

$$f = \frac{V_{1a}}{V_{2a}} = \frac{3r + R_R}{2r + R_R} = \frac{I_{1\_a\_via}}{I_{2\_a\_via}}$$

(1)

The lateral voltage gradient, defined as $l=V_{1a}/V_{5a}$ is computed using resistances approximated by shortest paths shown in Figure 4(b).

There are two shortest length path_5 and path_1, and only one passes through via *5*. Thus the resistance difference between path_5 and path_1 is *r*/2. Therefore, we have

$$l = \frac{V_{1a}}{V_{5a}} = \frac{R_{total\_path}}{R_{total\_path} - r/2} = \frac{R_L + R_B}{R_L + R_B - r/2}$$

(2)

If horizontal and vertical wire segments are not of equal length, let $L_R=L_L+\Delta L_1$ and $L_B=L_T+\Delta L_2$; we can scale the wires such that horizontal and vertical segments are of equal length by replacing voltage $V_b$ and $V_c$ using

$$V_{b\_eq} = V_b \pm \frac{\rho \cdot \Delta L_1}{S_2} I_b, \quad V_{c\_eq} = V_c \pm \frac{\rho \cdot \Delta L_2}{S_1} I_c$$

(3)

In equation (3), $S_i$ denotes the cross sectional area of a wire on metal layer *i*, $\rho$ is the resistivity, $I_b$ is the current in the right wire segment when $V_b$ is applied, $I_c$ is the current in the bottom wire segment when $V_c$ is applied. The correcting terms have a positive or negative signs depending on current directions.

Based on above analysis, we can express all via currents first in terms of $I_{1\_a\_via}$ and value of $I_{1\_a\_via}$ can be derived using forward and lateral voltage grandniece.

Experimental results for 4x4 arrays indicate that the maximal inaccuracy in computing via currents is less than 1% of the total current value. This error is contributed mostly by the inaccuracy of forward gradient computation. Our methods can be extended to *N* x *N* via array by modifying equivalent circuit and shortest paths. We note that the analysis developed in this Section also provides an explanation why current distributes unevenly in multi-vias.

## 3. EM Failure Sequence

### A. Basic Assumptions

It should be noted that each time when a via fails EM, it causes current redistribution in the via array. [5] To simplify our analysis, we assume the current redistribution happens instantly. This is a mild assumption since current redistribution usually happens and stabilizes in a short amount of time.

With the above assumption, we can describe the via failure sequence by repeating the following steps: a via fails→remove it→current redistributes→a new via fails. Figure 6 shows a sample via array failure sequence for a 2x2 via array.

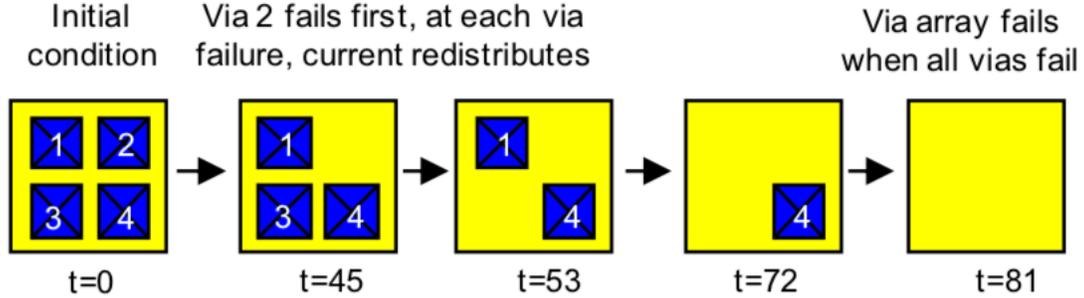

**Figure 6. Sample via failure sequence**

### B. Single Via EM Failure

Researches and studies have been well conducted for single via failure analysis. The random microstructure of copper body/surface contributes significantly to the variation of via failure time. Lognormal distribution can be used to accurately describe a single via failure as shown in equation (4), where $t_{50}$ is the mean time to failure (MTTF), $\sigma$ is the shape factor.

$$F(t;t_{50},\sigma) = \int_0^t \frac{1}{\sqrt{2\pi}\sigma t} e^{-(\ln t - \ln t_{50})^2/2\sigma^2}$$

(4)

Values of $t_{50stress}$ and $\sigma_{stress}$ can be extracted from measurements. In typical EM failure analysis, the $t_{50}$ value is a function of the current, while $\sigma$ is assumed to be constant for a specific technology. Black's equation [6] is used to model the dependence of $t_{50}$ on current density:

$$t_{50} = Aj^{-n} e^{\left(\frac{E_a}{kT}\right)}$$

(5)

In Equation (5), $A$ is an experimental constant, $j$ is the current density, $n$ is a scaling factor, $E_a$ is the activation energy, $k$ is Boltzmann constant, and $T$ is the absolute temperature. In general, it is believed that for void nucleation $n=2$, for void growth $n=1$, and with Joule heating $n > 2$. Most of the EM failures involve both void nucleation and growth. In our work, we assume $n=2$. With a reference $t_{50stress}$, other $t_{50}$ values for different current densities can be easily determined, and $\sigma$ is always equal to $\sigma_{stress}$.

## C. Via Array EM Failure

The memory effect of previous EM stress is the most difficult part to deal with for deriving a via array lifetime [7]. Also, the later via failure depends on previous via failures. We propose a via failure sequence $V_f$ to describe the whole EM degradation process.

**Definition:** $V_f = [(t_1,k_1),(t_2,k_2),\ldots,(t_N,k_N)]$, is a sequence that records via failure time $t$ and index $k$ of a failing via; $N$ is the total number of vias.

A sample via failure sequence corresponding to Figure 9 is [(45,2), (53,3), (72,4), (81,1)]. The time and via index are random numbers, therefore, theoretically we need to traverse all possible failure sequences and compute distributions for all $t$ and $k$. This is impractical due to exponential search space implied. Our goal is to determine only the distribution of the *last* via failure time $t_N$ and the intermediate distributions are of no interest. We use a Monte Carlo approximation to sample intermediate $t$ and $k$ instead of considering all possible values. In this way, as long as the previous via failure sequence is determined, the next via failure time and index number are easy to calculate. This method has a linear complexity on $N$.

Knowing the previous via failure sequence, stress time translation is used to account for the memory effect. The translation rule is given by Equation (6).

$$\left(\frac{i_{m-1}}{i_m}\right)^n = \frac{t_{m-1}'}{t_{m-1}}$$

(6)

In Equation (9), $n$ is the same exponent in Equation (5) ($n$=2 in our analysis); $i_{m-1}$ and $i_m$ are previous and present currents on a via; $t_{m-1}$ is the previous via failure time; $t_{m-1}'$ is the translated stress time. For example, in Figure 6, assume via *3* carries a current density of 10mA/mm² from $t$=0 to 45s, and after via *2* fails, it carries current density of 15mA/mm². The stress of 10mA/mm² for 45s can be translated to an equivalent stress of 15mA/mm² for 20s.

Now, given the condition that via *3* does not fail under 15mA/mm2 for 20s, the conditional CDF of via *3* is given by Equation (7), where $k$ is the via index, in this particular case, $k$=3 and $t_{m-1}'$=20s.

$$F_k'(t) = \frac{F_k(t + t_{m-1}') - F_k(t_{m-1}')}{1 - F_k(t_{m-1}')}$$

(7)

Since vias are connected in parallel, the weakest via is the next one to fail. The conditional via failure CDF is then combined to determine the next via using Equation (8), and the probability of next via failure to be via $k$ is given in Equation (9).

$$F_{next}(t) = 1 - \prod_{k \in V_g} (1 - F_k'(t - t_{m-1}))$$

(8)

$$P_{next=k} = \int_{t=0}^{\infty} \left[ f_k'(t) \cdot \prod_{l \neq k, l \in V_g} (1 - F_l'(t)) \right] dt$$

(9)

$V_g$ denotes the set of vias that are still conducting and $f_k'(t)$ is the probability density function (PDF) for $F_k'(t)$. With the above equations, a Monte Carlo approximation can sample $t$ and $k$ accordingly at each via failure step and generate the via failure sequence. The averaged results from multiple Monte Carlo runs are used to approximate the overall via array CDF.

# 4. Experimental Results

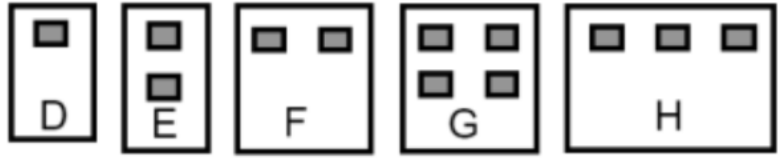

**Figure 7. Via array layout**

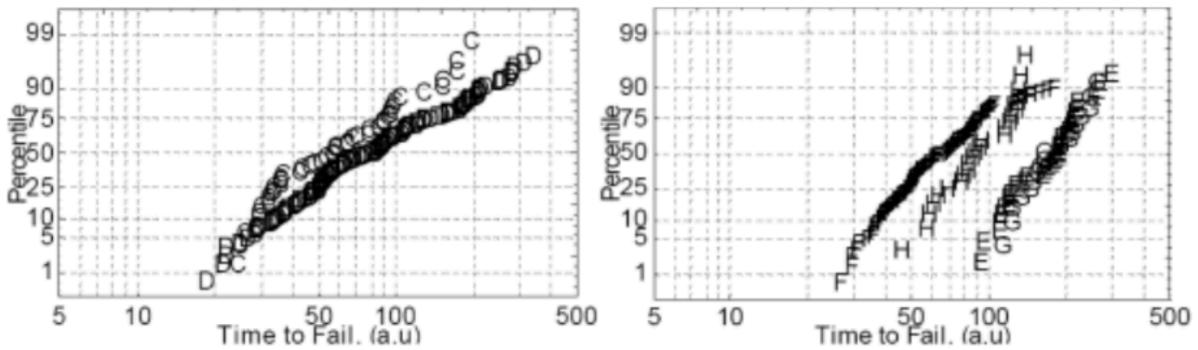

**Figure 8. Via array EM failure test results**

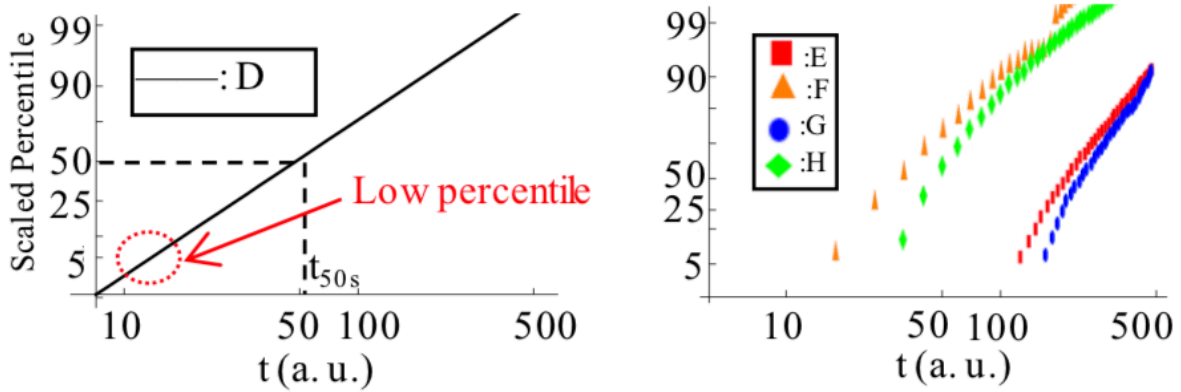

**Figure 9. Via array EM failure analysis results**

Our proposed EM via array failure analysis can be verified through simulating a few via arrays. There are papers providing real measurement results, therefore can be used as reference. The via array structures are shown in Figure 7. The nominal via size is 0.14mm x 0.14mm and the wire thickness is 0.19mm. Test stress is at 300°C under 25mA/mm². Figure 8 shows EM test results from [8], while Figure 9 shows our analysis results.

From above figures we can observe that the real stress test data matches well with the failure distribution obtained from simulation results. Note that case D is a single via, therefore a simple straight line analytical solution can be derived, also low percentile part is of more interest from reliability of view; solutions for other cases are from Monte Carlo approximation [9]. It is observed that for some cases (such as F) $t_{50}$ can be smaller than the single via case, however because of the small $\sigma$, the lower percentile reliability for case F still wins.

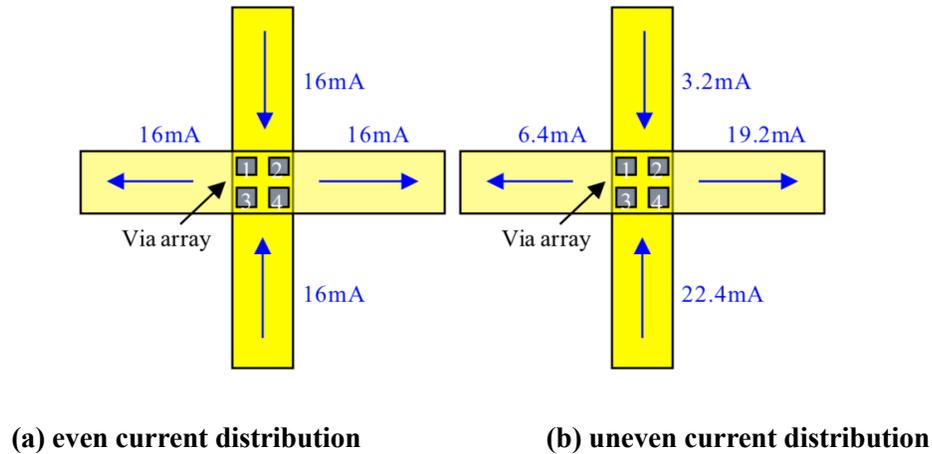

**(a) even current distribution**         **(b) uneven current distribution**

**Figure 10. 4x4 multi-via test examples**

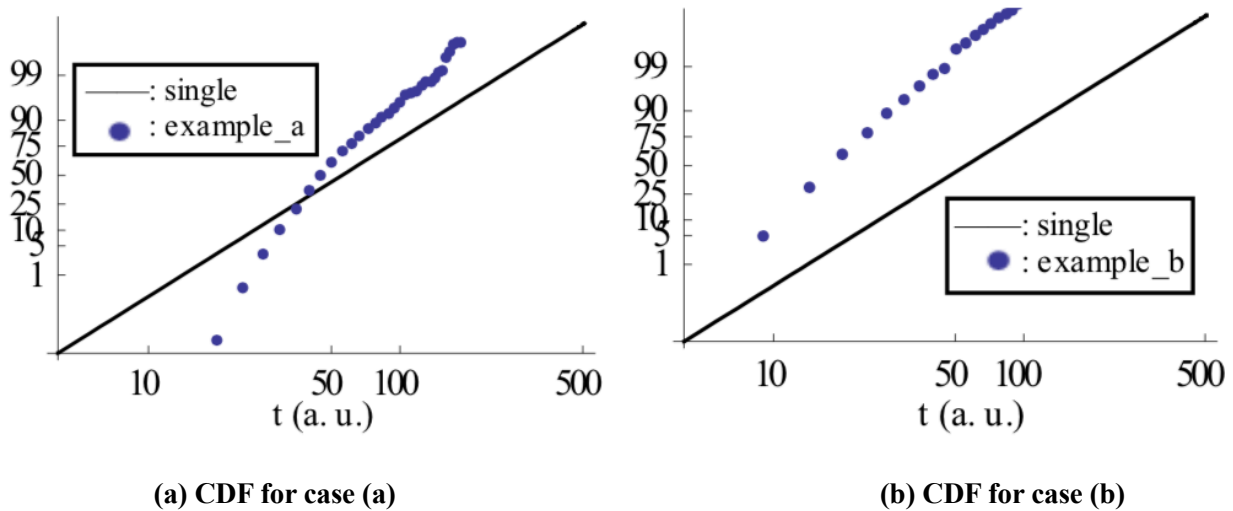

**(a) CDF for case (a)**         **(b) CDF for case (b)**

**Figure 11. CDF comparison**

Our analysis is further applied to some typical power grid via array structures. The via structures are constructed based on the IBM power grid benchmarks [10], with current values based on benchmark solutions. To generate the vias we use the following additional industrial geometric parameters: wire thickness is 0.6mm, wire width is 2mm, via size is 0.8mm x 0.8mm. We assume 10mA/mm² is the EM current limit for a reference single via with $t_{50ref}$. We consider two cases shown in Figure 10: case(a) all wires carry the same current density of 13.12mA/mm² of current, leading to an evenly distributed via current; case (b) currents in the wires are unequal at 3.2mA/mm², 6.4mA/mm², 19.2mA/mm² and 22.4mA/mm², thus the via array has highly uneven current distribution. The total current passing through the via array is 25.6mA.

Traditional EM rules assume an even current distribution [11], so for case (a), each via carries 10.25mA/mm$^2$, which violates the EM current limit; for case (b), each via carries 10mA/mm$^2$, and is within the EM current limit. However, considering just the average currents is not sufficient since we have and uneven current distribution and the multi-via redundancy effect acting together to determine the via array EM reliability. The initial current distributions for case (b) are 7.36mA/mm$^2$, 10.08mA/mm$^2$, 10.24mA/mm$^2$ and 12.32mA/mm$^2$.

In Figure 11, the via array lifetime distributions for cases (a) and (b) are plotted against the reference of a single via lifetime.

The results are counter-intuitive and do not agree with the traditional EM reliability evaluation [12]. For case (a), $t_{50}$ is close to $t_{50ref}$, and the small $\sigma$ makes it an EM reliable design at lower failure percentiles. On the other hand, for case (b), $t_{50}$ is obviously worse than $t_{50ref}$. For lower percentiles it appears better than the reference single via due to a relatively small $\sigma$, but it is very close to the reliability boundary and should be considered EM-unsafe. These two examples contradict common belief and demonstrate a need for a proper method to evaluate multi-via EM reliability as discussed in this paper.

## 5. Conclusions

Multi-vias are widely used to connect wires from different layers to improve EM reliability, but no detailed via array lifetime evaluation methods exist. In this paper, we demonstrated that current distributes unevenly in multi-vias and we explained why. We developed a fast model to calculate currents flowing through individual vias, and proposed a step-by-step multi-via failure model. Each time a via fails, current redistribution is calculated and the via memory effect is accounted for using stress time translation. We apply a Monte Carlo approximation to generate via failure sequences and the overall via array lifetime distribution. Experimental results show that our predicted lifetimes correlate well with EM stress test results. For multi-vias, both the redundancy effects and uneven current distributions affect reliability, leading to counter-intuitive results in real circuit when the current distribution is high.